\documentclass[%
 reprint,
 amsmath,amssymb,
 aps,
]{revtex4-2}

\usepackage{graphicx}
\usepackage{dcolumn}
\usepackage{bm}
\usepackage{upgreek}
\usepackage{xcolor}

\begin{document}

\preprint{APS/123-QED}

\title{Hybrid photonic-plasmonic cavity design for very large Purcell factors at telecom wavelengths}

\author{Angela Barreda}
\affiliation{Friedrich Schiller University Jena, Institute of Solid State Physics, Helmholtzweg 3, 07743 Jena, Germany}
\affiliation{Friedrich Schiller University Jena, Institute of Applied Physics, Abbe Center of Photonics, Albert-Einstein-Str. 15, 07745 Jena, Germany}

\author{Laura Mercad\'{e}}%
\affiliation{Nanophotonics Technology Center, Universitat Polit\`{e}cnica de Val\`{e}ncia, Camino de Vera s/n, 46022, Valencia, Spain}%
\affiliation{Departament d’Enginyeria Electr\'{o}nica i Biom\'{e}dica, Facultat de F\'{i}sica, Universitat de Barcelona.}

\author{Mario Zapata-Herrera}%
\affiliation{Materials Physics Center CSIC-UPV/EHU, 20018 Donostia-San Sebastian, Spain}

\author{Javier Aizpurua}%
\affiliation{Materials Physics Center CSIC-UPV/EHU, 20018 Donostia-San Sebastian, Spain}
\affiliation{Donostia International Physics Center DIPC, 20018 Donostia-San Sebastian, Spain}
\author{Alejandro Mart\'{i}nez}%
\affiliation{Nanophotonics Technology Center, Universitat Polit\`{e}cnica de Val\`{e}ncia, Camino de Vera s/n, 46022, Valencia, Spain}%

\date{\today}

\begin{abstract}
Hybrid photonic-plasmonic cavities can be tailored to display  high $Q$-factors and extremely small mode volumes simultaneously, which results in large values of the Purcell factor, $F_\mathrm{P}$. Amongst the different hybrid configurations, those based on a nanoparticle-on-a-mirror (NPoM) plasmonic cavity provide one of the lowest mode volumes, though so far their operation has been constrained to wavelengths below 1 $\upmu$m. Here, we propose a hybrid configuration consisting of a silicon photonic crystal cavity with a slot at its center in which a gold nanoparticle is introduced. This hybrid system operates at telecom wavelengths and provides high $Q$-factor values ($Q \approx 10^{5}$) and small normalized mode volumes ($V_\mathrm{m} \approx 10^{-4} $), leading to extremely large Purcell factor values, $F_{\mathrm{P}} \approx 10^{7}-10^8$. The proposed cavity could be used in different applications such as molecular optomechanics, bio- and chemo-sensing, all-optical signal processing or enhanced Raman spectroscopy in the relevant telecom wavelength regime.
\end{abstract}

\maketitle


\section{\label{sec:level1}Introduction \textbackslash\textbackslash}

The local density of optical states (LDOS) quantifies the available amount of electromagnetic states that can be occupied by a photon in a certain position of a system. A high local value of the LDOS will result in enhanced light-matter interaction there. This way, the environment of an emitter can be tailored to maximize the LDOS, for instance, using optical cavities. At resonance, the LDOS gives us the Purcell factor, $F_\mathrm{P}$, which originally accounted for the enhancement of the spontaneous emission of an atom in a cavity \cite{Purcell1946,Koenderink2010,Zambrana2015}. By definition, the Purcell factor is proportional to the ratio between the $Q$-factor and the mode volume $V_\mathrm{m}$ as \cite{novotny,sprik_1996}: 

\begin{equation}
\label{eq_pur}
F_{\mathrm{P}}=\frac{3}{4\pi^2}\frac{Q}{V_\mathrm{m}}  
\end{equation}  

where $V_\mathrm{m}$ is the mode volume normalized by the wavelength ($\lambda$) over the local refractive index ($n$) cubed $(\lambda/n)^3$, and $Q$ is the quality factor. 
This suggests a route to design photonic systems that exhibit simultaneously high $Q$-factor and ultra small $V_\mathrm{m}$ to maximize  $F_\mathrm{P}$ and enhance light-matter interaction. 

Dielectric cavities can greatly enhance the Purcell factor since they can have high $Q$-factor values as a result of the negligible absorption loss \cite{choy}. However, the diffraction limit usually prevents them to achieve subwavelength-scale mode volumes, so $V_\mathrm{m} \approx 1$ unless special ``slotted'' configurations are used \cite{PRL_cavity,science_cavity,Seidler}. On the contrary, plasmonic cavities formed by metallic nanoparticles (NPs) can overcome the diffraction limit leading to ultrasmall $V_\mathrm{m}$ values\cite{Maier:2007}. However, Joule's losses in the metal reduces the reachable $Q$-factor to values $Q \approx 10$ \cite{Doeleman2016,Akselrod}.

In the last few years, hybrid photonic-plasmonic cavities have been introduced as novel systems providing large $F_\mathrm{p}$ as they can potentially combine the best of both worlds: high $Q$-factor values due to the dielectric photonic cavity and small mode volumes thanks to the metallic NP \cite{Doeleman2016,Mukherjee11,PhysRevA.95.013846,Mukherjee12,Dezfouli,Palstra2019,Doeleman2020,Zhang:20,Ehsan,Hong,Klusmann}. Indeed, by modifying the coupling between photonic and plasmonic modes, the $Q$-factor and $V_\mathrm{m}$ can be tuned between the values of the bare NP and the cavity \cite{Palstra2019}. Different configurations of hybrid photonic-plasmonic cavities have been suggested so far. Some of the most attractive ones consist of a bow-tie nanoantenna placed on a dielectric photonic crystal nanobeam cavity operating at transverse-electric polarization \cite{Palstra2019} ($F_\mathrm{p} \approx 1 \cdot 10^{4}$). The main limitation of this configuration is that the plasmonic gap has to be defined lithographically reaching minimum values of several nm, which prevents reaching the nm-scale (and below) gaps that leads to extreme plasmonic localization \cite{Hatab2010,Crozier,Duan_2012}. 

Nanoparticle-on-a-mirror (NPoM) cavities are far superior to achieve the smallest possible mode volumes, as plasmonic gaps
around 1 nm and below can be created. The reason is that NPs can be deposited on top of self-assembled monolayers (SAMs) formed on metallic surfaces so the plasmonic gap widths equal the SAM thickness \cite{Leveque2006,Ciraci2012,Baumberg2019,Carnegie2018,Barbry}. Recently, a hybrid system combining a NPoM configuration with GaP photonic crystal cavities - operating under transverse-magnetic polarization at visible wavelengths - has been introduced \cite{Barreda:21}. Thanks to the NPoM-based design, nm- and sub-nm-scale plasmonic gaps potentially can be achieved and combined with relatively large $Q$-factors. By means of this novel hybrid system, the Purcell factor values are enhanced one order of magnitude with respect to the hybrid configurations based on bow-tie nanoantennas ($F_\mathrm{p} \approx 1 \cdot 10^{5}$) \cite{Barreda:21}. 

Most of the previous works on hybrid photonic cavities have been focused on the visible wavelength range due to the huge interest in enhancing light-matter interaction in that band \cite{Palstra2019}. However, at the telecom wavelengths ($\lambda \approx 1550$ nm) is highly interesting. First, because of the multiple available applications (mostly related to high-speed optical communications and data processing) as well as advanced available equipment (laser, amplifiers and high-speed photodetectors); and, second, because of the emergence of the silicon photonics area, which has become the mainstream technology for photonic integrated circuits. Indeed, there have been many demonstrations of silicon cavities operating at telecom wavelengths with $Q$ factors over $10^{4}$ using lithographically-defined 1D ($F_\mathrm{p} \approx 2.7 \cdot 10^{6}$) \cite{apl} or 2D ($F_\mathrm{p} \approx 1.7 \cdot 10^{5}$) \cite{nature} photonic crystal cavities whilst showing $V_\mathrm{m} \approx 1 $. Reaching $V_\mathrm{m} < 1 $ becomes feasible by defining thin slots in silicon and taking advantage of the discontinuities of the normal electric field \cite{Robinson,PRL_cavity}, but with the minimum slot size imited by lithography. In previous work \cite{science_cavity}, complex subwavelength structuring of silicon (down to $\approx 10$ nm) enabled to get also $V_\mathrm{m} < 2 \cdot 10^{-4}$. 

Owing to this technological interest, hybrid systems have also been recently proposed in silicon technology for telecom wavelengths. In \cite{Zhang:20}, a hybrid photonic-plasmonic nano-cavity, constituted by an L3 photonic crystal nano-cavity and a plasmonic bow-tie nanoantenna was shown to have an ultrahigh figure of merit $Q/V$ of $8.4 \times 10^6$ $(\lambda/n)^{-3}$ ($F_\mathrm{p} \approx 6.3 \cdot 10^{5}$). However, either in the only-silicon cavity \cite{science_cavity}, or in the hybrid approach \cite{Zhang:20} the extreme localization enabled by the NPoM is still missing. 


In this work, we propose a hybrid photonic-plasmonic cavity that combines silicon and the NPoM configuration to get extreme field localization at resonance. This hybrid consists of a silicon photonic cavity with a slot at its center \cite{Robinson,Seidler}, operating at transverse-electric polarization at telecom wavelengths, and a gold NP, which is located at the slot's cavity. The NP can be potentially separated from the silicon walls of the slot by a SAM, thus reaching the nm- and even sub-nm-scale localized fields at resonance. Our design achieves $Q \approx 1\cdot10^{5}$ and $V_\mathrm{m} \approx 1\cdot10^{-4} $, giving rise to Purcell factor values of $F_\mathrm{P} \approx{1\cdot10^7}-1\cdot10^8$, which are two-three orders of magnitude larger than those reported previously based on NPoM cavities combined with photonic crystals and two orders of magnitude larger than those corresponding to hybrid photonic-plasmonic cavities operating at telecom wavelengths. This is mainly due to the nanoscale confinement provided by the slot-nanoparticle combination and the radiation suppression resulting from the photonic crystal structure. 

We also suggest a route to fabricate - combining top-down and bottom-up approaches - the hybrid cavity, which may be used to improve the performance of current silicon photonic devices as well as to envisage new applications in the relevant telecom-wavelength domain.


\section{Hybrid system description }
\begin{figure} 
\includegraphics[width=\columnwidth]{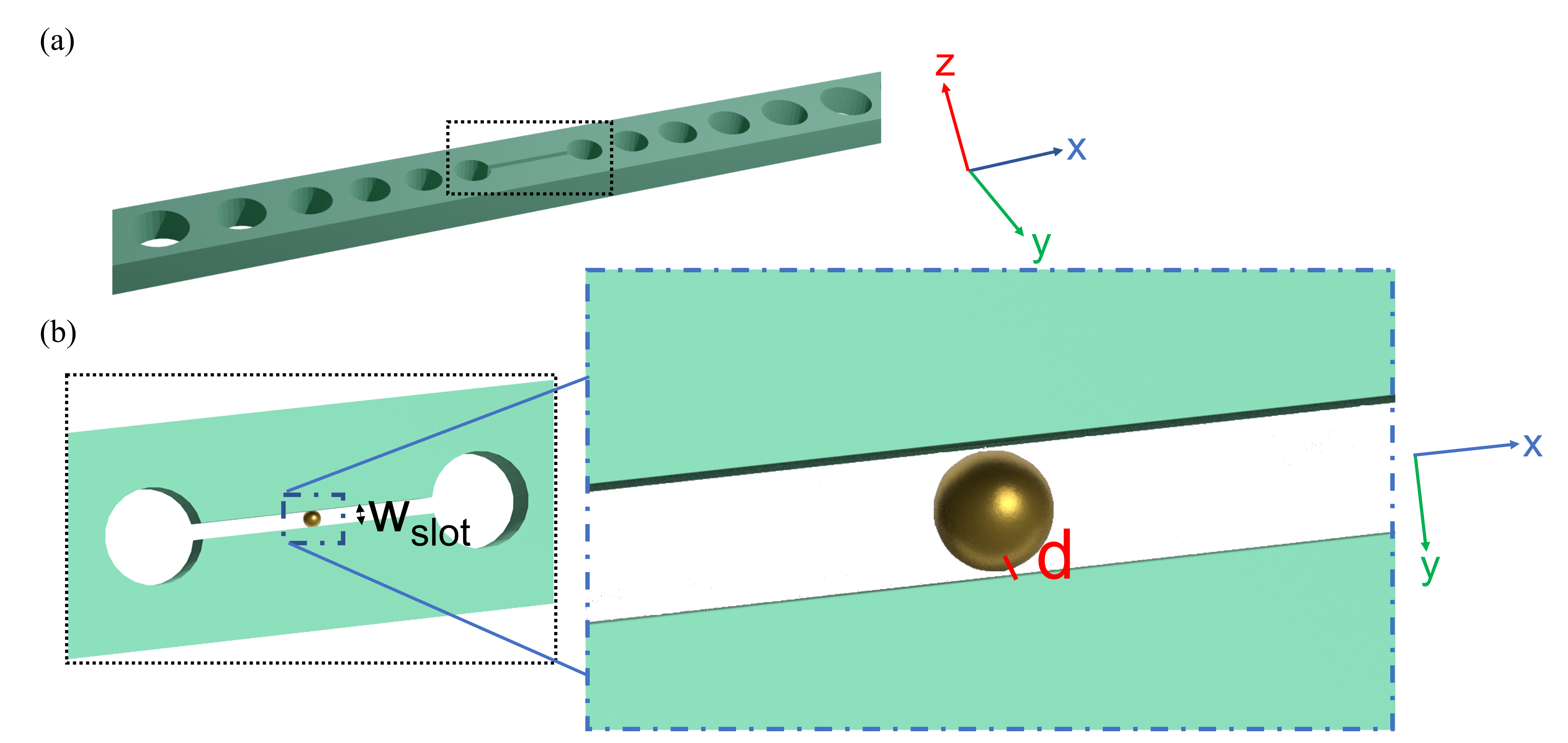}
\caption{Scheme of the proposed hybrid system. (a) Geometry of the 1D slotted photonic crystal cavity. (b) Top view of the cavity's center, where the slot of width $W$ can be distinguished. A gold spherical NP is located inside the slot. The distance between the NP and the wall of the cavity in the $y$-direction is $d = 1$ nm as shown in the close view marked with dashed-dotted line.}
\label{fig_geo}
\end{figure}

Figure \ref{fig_geo} depicts the proposed hybrid photonic-plasmonic configuration. A gold NP is located at the center of the slot of a 1D silicon photonic crystal cavity operating under transverse-electric polarization at telecom wavelengths. The distance between the NP and the photonic crystal wall (gap, $d$) is $d = 1$ nm, which approximately models the thickness of a SAM. The silicon wall behaves here as the mirror, though with lower reflectance than for a metallic one. 
Such a small gap, which is not defined lithographically but after depositing the NP after the silicon cavity is fabricated, should allow for extreme light confinement, thus giving rise to small $V_\mathrm{m}$. In addition, high $Q$-factors can be achieved due to the photonic bandgap of the dielectric cavity. 

\section{Methods} 


The eigenfrequencies and LDOS calculations shown in this work were obtained by means of the Finite Element Method (FEM), implemented in the commercial software COMSOL Multiphysics \cite{Comsol}. In particular, the Radio Frequency (RF) Module was used. The optical constants for the gold NP were taken from Ref. \cite{Palik1998}. The refractive index for silicon was considered as $n = 3.48$. To attain the LDOS, the structure was illuminated by an electric point-dipole source, whose dipole moment was considered along the $y$-axis (i.e. along the axis connecting the sphere and the cavity). The normalized LDOS is defined as 

\begin{equation} 
LDOS = \frac{P_\mathrm{rad}+P_\mathrm{no-rad}}{P_\mathrm{rad\ vacuum}}
\label{eq_ldos}
\end{equation}
                   
where $P_\mathrm{rad}$ is the radiative power density and $P_\mathrm{no-rad}$ is the total power loss density. The numerator corresponds to the radiative and nonradiative power emitted by the dipole coupled to the structure. However, the denominator contains the radiative power emitted by the dipole in a vacuum. The dipole was located at the center of the structure and in the middle of the gap between the NP and the cavity. Experimentally, this spacer would be filled by a SAM. The hybrid geometry was surrounded by a cylindrical air region of radius 9 $\mu$m. An additional smaller cylinder with a radius of 5 $\mu$m and made of air was placed in the center of the larger cylinder. The scattered power ($P_\mathrm{rad}$) was calculated at the boundaries of the smaller cylinder, whereas the total power loss density was obtained by means of the volume integration of the losses in the metallic NP. The dipole source was surrounded by a sphere with a diameter that is equal to the gap size ($d = 1$ nm) to ensure a sufficient fine grid in close proximity to the dipole source. The mesh of the surrounding air medium was chosen to be smaller than 600 nm. The mesh of the gold sphere was smaller than 6 nm. The chosen sizes for the mesh of the slot, cavity surrounding the slot, holes and cavity surrounding the holes were smaller than 15 nm, 22.5 nm, 45 nm and 52.5 nm, respectively. The mesh of the sphere surrounding the dipole source was smaller than 0.2 nm. The cylindrical region of air was surrounded by a perfectly matched layer (PML) with a thickness of 3 $\mu$m. 

The near-field plots at the eigenfrequencies of interest were attained with the eigenmode solver of COMSOL Multiphysics. The same mesh was used for the LDOS and eigenmode calculations and for the different analyzed structures: dielectric (no NP) and hybrid cavity. In all the cases, the same geometry was considered, and only optical constants of the materials (if present or not) were changed appropriately. 

\section{Results} 

\subsection{Photonic band diagram} 
\begin{figure}[h!]
\includegraphics[width=\columnwidth]{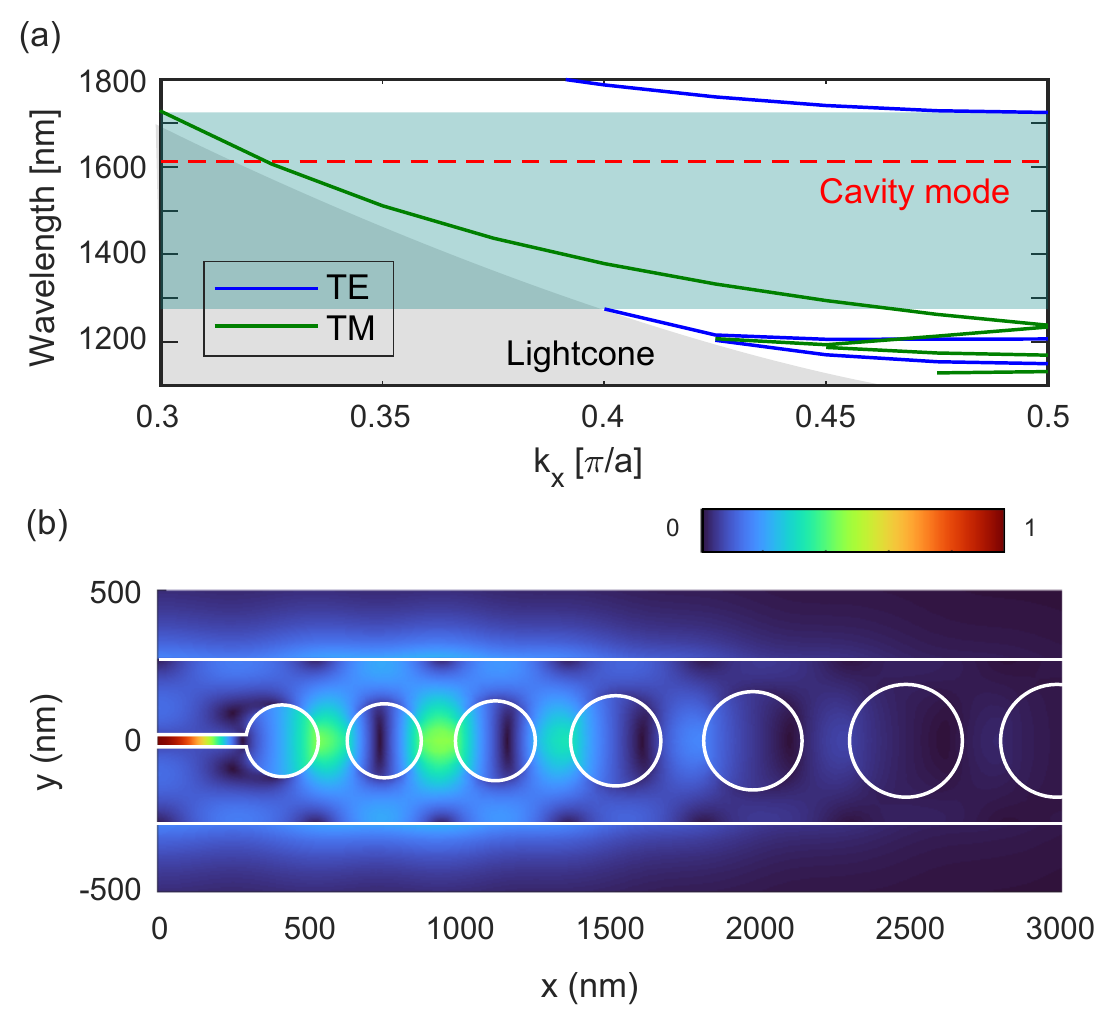}
\caption{a) Photonic band diagram of the mirror unit cell. In blue (green) the TE-like (TM-like) bands are depicted. The shadowed region corresponds to the TE-like quasi bandgap where the photonic crystal cavity mode confined in the middle region is located (depicted with a red dashed line). b) Mode profile of $|\mathrm{E}|$ (X-Y crosscut) normalized to its maximum value for the slotted photonic crystal without the spherical gold nanoparticle.}
\label{fig_band}
\end{figure}

The 1D silicon photonic crystal cavity consists of a silicon waveguide of width $h = 550$ nm and thickness $t = 220$ nm patterned with holes and with a slot in the defect section, as depicted in Figure \ref{fig_geo}(a). The beam is drilled by a set of circular holes to produce a transverse-electric (TE) band gap around $\lambda = 1550$ nm. Figure \ref{fig_band}(a) shows the TE-like band diagram when the holes are spaced by a distance $a = 580$ nm and have a radius $r = 0.365a$, showing a wide bandgap (shadowed region) as expected. To build the cavity, we include two photonic crystal mirrors with the previous dimensions at each side of the cavity whilst the period is adiabatically reduced from $a = 580$ nm down to $a = 331.5$ nm (keeping constant the $r/a$ ratio) when moving to the cavity center. 
This permits to blue-shift the dielectric band and get a confined mode in the band gap, as depicted by the red dashed line in Fig. \ref{fig_band}(a). Finally, a slot section of width $w = 40$ nm and length $l = 547$ nm is introduced between the central holes of the cavity to enhanced the transverse electric field. Notice that this slot width is achievable using standard silicon nanofabrication \cite{Seidler}. 


Figure \ref{fig_band}(b) represents  the near-field map at the eigenfrequency of the confined mode ($\lambda = 1613$ nm) in the 1D photonic crystal cavity following the previous design. It can be clearly seen that the electric field is well confined in the slot placed at the center of the cavity. It is worth mentioning that the cavity resonance is red-shifted with respect to the gold NP resonance. In previous works, it was evidenced that this condition is necessary to improve the performance of the hybrid cavity with respect to the bare components \cite{Palstra2019, Doeleman2016, Doeleman2020}. 



\subsection{Comparison $Q$-$V_{\mathrm{m}}$ values for the bare cavity and hybrid system } 

\begin{figure}[h!] 
\centering 
\includegraphics[width=\columnwidth]{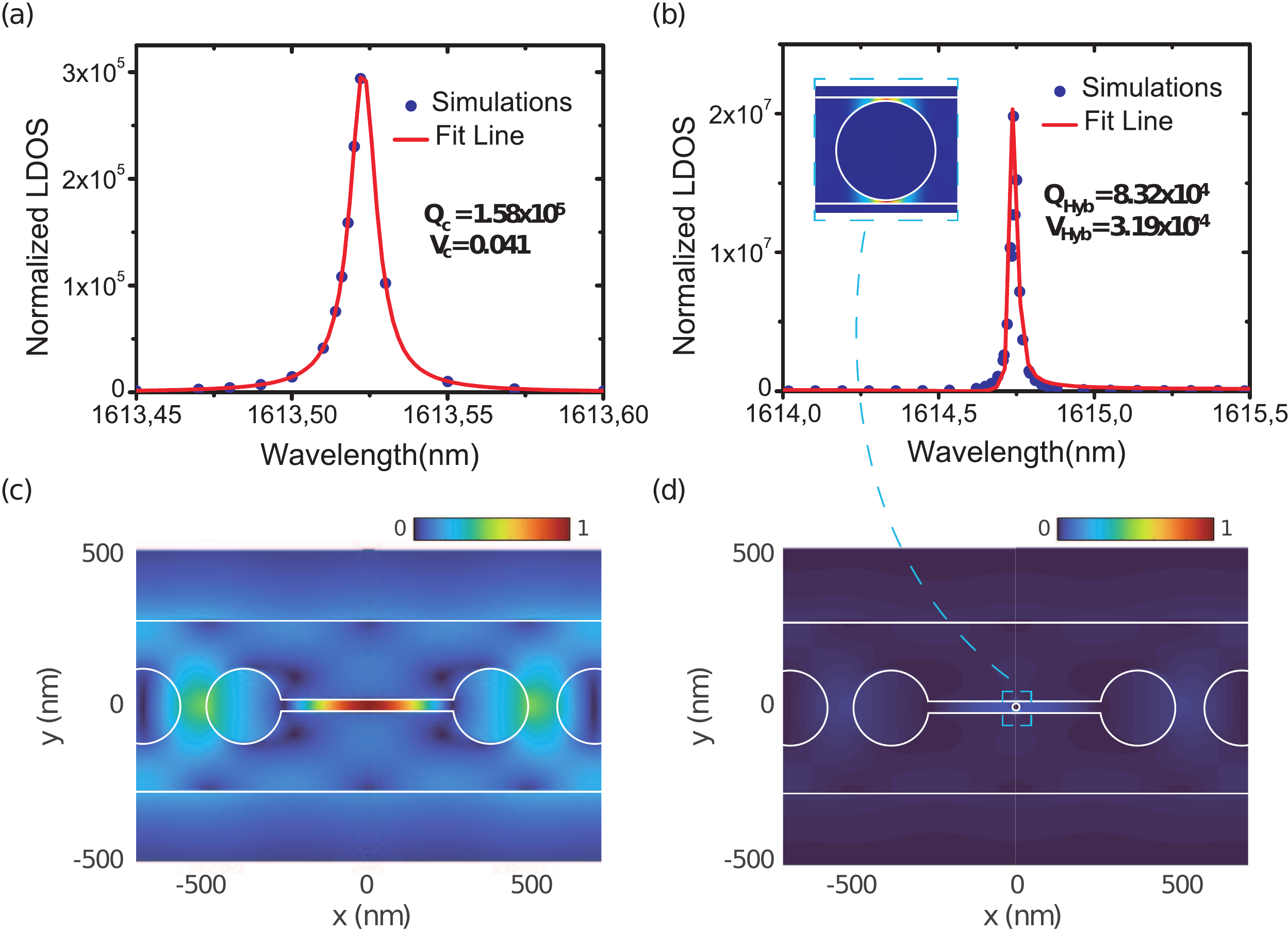}
\caption{Normalized LDOS, $Q$-factor and $V_\mathrm{m}$ for a) the bare silicon photonic crystal ($Q_{\mathrm{c}}$, $V_{\mathrm{c}}$) and b) the hybrid plasmonic-photonic cavity ($Q_{\mathrm{Hyb}}$, $V_{\mathrm{Hyb}}$). The slot size is $w = 40$ nm and the radius of the gold NP is $R = 19$ nm. Mode profile of $|\mathrm{E}|$ (X-Y crosscut) normalized to its maximum value in the slot for c) the bare silicon photonic crystal and d) the hybrid photonic-plasmonic cavity. The inset figure in panel b) shows a close view of the confined mode profile $|\mathrm{E}|$ of the hybrid photonic-plasmonic cavity in the cavity's center.}
\label{fig_slot_40}
\end{figure}

In Figs. \ref{fig_slot_40}(a) and (b), we represent the calculated LDOS for the bare cavity (slot width $w = 40$ nm) and the hybrid (Hyb) system composed of the silicon cavity plus a gold NP of radius $R = 19$ nm, respectively. For the bare cavity, the $Q$-factor is retrieved from the fitting of LDOS to a Lorentzian function. Once the $Q$-factor is known, the mode volume is retrieved by the inversion of Eq. \ref{eq_pur}. The obtained $Q$-factor of the bare cavity is $Q_{\mathrm{c}} = 1.6\cdot10^{5}$, whilst the mode volume value is $V_\mathrm{c} = 4\cdot10^{-2}$. As expected, high $Q$ values are achieved, and could even be higher by a more meticulous design of the bare cavity. However, the mode volume is still large in comparison to the values that can be obtained by considering metallic nanostructures ($V_\mathrm{NP} \approx 1\cdot10^{-6}$). When the NP is introduced in the gap, the $Q$-factor of the hybrid system is only slightly diminished ($Q_{\mathrm{Hyb}} = 8.3\cdot10^{4} $) but there is a two orders of magnitude reduction of $V_\mathrm{m}$ down to $V_{\mathrm{Hyb}} = 3.2\cdot10^{-4}$ as a result of the hybridization of the metallic NP with the photonic cavity.  For the hybrid cavity a Fano fitting is necessary due to the hybridization of two different modes: the corresponding to the cavity and that of the NP.

If we observe the near-field maps of the electric field in the slot for the bare cavity and the hybrid system (see Fig. \ref{fig_slot_40}(c) and (d), respectively), we see that the gold NP squeezes the slot field into a nm-scale region, which was the purpose of introducing the NPoM approach. Therefore, the hybrid system provides a way for extreme localization of telecom-wavelength fields whilst keep large $Q$ values. 



\newpage 

\subsection{Evolution of $Q$-$V_\mathrm{m}$ values with the slot size} 

\begin{figure}[h!]
\includegraphics[width=\columnwidth]{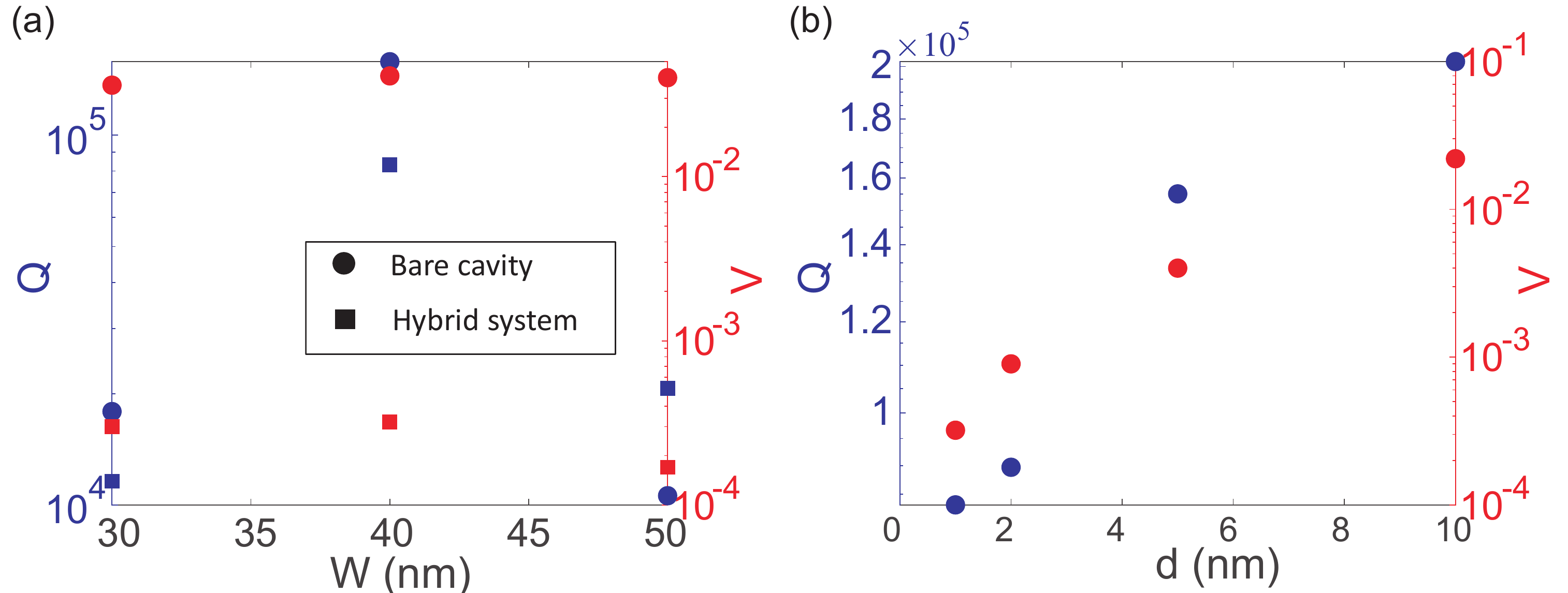}
\caption{(a) $Q$-factor (blue markers) and $V_{\mathrm{m}}$ (red markers) values for the bare cavity (points) and hybrid system (squares) as a function of the slot size $W$. The radius of the gold NP has been varied accordingly to keep the gap distance between the NP and the wall of the photonic crystal to $d = 1$ nm. (b) $Q$-factor (blue markers) and $V_{\mathrm{m}}$ (red markers) values for the hybrid system as a function of the gap size. The size of the slot was kept constant to $W = 40$ nm. The $y$-axis is represented in logarithmic scale.}
\label{fig_slot_size}
\end{figure}

From an experimental point of view, some differences in the size of the slot can be expected with respect to the nominal parameters due to imperfections during the fabrication process. In this section, we analyze the influence of the slot width in the $Q$-factor and $V_{\mathrm{m}}$ values. Figure \ref{fig_slot_size}(a) shows the $Q$-factor and $V_{\mathrm{m}}$ values as a function of the slot size for the bare cavity and the hybrid system. We observe that as the size of the slot is modified (both increased or decreased) with respect to the optimum value $w = 40$ nm, the $Q$-factor decreases almost one order of magnitude. However, $V_{\mathrm{m}}$ values are not so affected due to changes in $W$, especially for the bare cavity. In all the cases, the gap distance between the gold NP and the wall of the cavity is $d = 1$ nm, which means that the NP radius is modified accordingly for each $w$ value. 


\newpage

\subsection{Evolution of $Q$-$V_\mathrm{m}$ values with the NP-cavity gap} 
Small imperfections in the chemical production of the NP can also generate imprecision in its size, provoking uncertainty in the gap distance between the NP and the wall of the photonic cavity. To analyze this effect, we have represented in Fig. \ref{fig_slot_size}(b) the $Q$-factor and $V_{\mathrm{m}}$ values for different gap sizes for a slot cavity of $W = 40$ nm. In particular, the size of the NP was changed between $R = 19$ nm and $R = 10$ nm, leading to variations in the gap from $d = 1$ nm to $d = 10$ nm. As the gap increases, both $Q_{\mathrm{Hyb}}$ and $V_\mathrm{Hyb}$ increase. This effect is due to the lower confinement of the electromagnetic radiation in the gap.

\section{Conclusions} 
In this work we have proposed a novel design of a hybrid photonic-plasmonic cavity, which combines a 1D silicon photonic cavity with a slot in its center, operating under transverse-electric polarization at telecom wavelengths, with a gold nanoparticle, located at the slot's center. This configuration shows how to merge the NPoM approach with silicon photonic crystal cavities in order to achieve extreme light confinement in nm-scale gaps. Through this nanostructure we have obtained $Q$-factor values for the hybrid system larger than $Q_{\mathrm{Hyb}} = 1\cdot10^{5}$ and mode volumes smaller than $V_{\mathrm{Hyb}} = 1\cdot10^{-4}$. This gives rise to Purcell factor values of $F_{\mathrm{P}} \approx 1\cdot10^7-1\cdot10^8$, which are two or three orders of magnitude higher than those obtained in previous NPoM hybrid photonic-plasmonic cavities \cite{Barreda:21} or metallic bow-ties antennas coupled photonic crystals operating under transverse-electric polarization \cite{Palstra2019}. 
If we extend the comparison to hybrid cavities operating at telecom wavelengths, we can conclude that our hybrid cavity provides two orders of magnitude larger than hybrid photonic-plasmonic nano-cavity reported in \cite{Zhang:20}. This is mainly due to the nanoscale confinement provided by the slot-nanoparticle combination and the radiation suppression resulting from the photonic crystal structure.

From a practical perspective, the 1D silicon photonic crystal cavity can be fabricated using standard silicon technology \cite{Seidler,science_cavity}. It is also possible to release the silicon beam from the substrate \cite{mercade} as in the configuration in this work. NPs can be later deposited by drop casting, potentially falling into the slot, as recently demonstrated for plasmonic nanoslits \cite{science_galland}. The use of optical forces could even be used to trap the NPs inside the slot, as demonstrated in slot waveguides \cite{nature_n}. Moreover, some techniques would enable the transfer of isolated gold NPs to the slot. If the beam is not released from the substrate, the photonic bandgap would be somewhat reduced but high-$Q$ modes are still possible and the NPs could be deposited on the silica substrate at the slot bottom. Thus, we foresee that a combination of top-down (for the photonic crystal cavity) and bottom-up (for the gold NPs deposition) techniques would make possible the experimental realization of this cavity, enabling molecular-scale light confinement together with large $Q$-factors.


Building hybrid cavities that operate at telecom wavelengths does not only offers new perspectives in applications already existing in that spectral regime (optical communications, photonic biosensing, nonlinear signal processing) but could also enable the transfer of (near-)visible-regime applications (such as surface-enhanced Raman spectroscopy \cite{LOS19-JSTQE} or optomechanically-driven frequency conversion \cite{science_galland,science_baumberg}) to a domain where a huge amount of high-quality instrumentation (tunable lasers, high speed photo-detectors, amplifiers) is available, whilst being implementable in silicon compatible chips.



\begin{acknowledgments}

A. B. thanks the financial support by the Deutsche Forschungsgemeinschaft (DFG, German Research Foundation) through the International Research Training Group (IRTG) 2675 ``Meta-ACTIVE'',  project number 437527638. L.M. thanks financial support from the Next generation EU program, Ministerio de Universidades (Gobierno de Espa\~na). A. M. acknowledges funding from H2020 European Commission (THOR-H2020-EU-829067); and Generalitat Valenciana (PROMETEO/2019/123). M. Z. H. and J. A. acknowledges Spanish Ministry of Science (Project No. PID2019-107432GB-I00) for financial support,

\end{acknowledgments}

\appendix



\begin{thebibliography}{41}%
\makeatletter
\providecommand \@ifxundefined [1]{%
 \@ifx{#1\undefined}
}%
\providecommand \@ifnum [1]{%
 \ifnum #1\expandafter \@firstoftwo
 \else \expandafter \@secondoftwo
 \fi
}%
\providecommand \@ifx [1]{%
 \ifx #1\expandafter \@firstoftwo
 \else \expandafter \@secondoftwo
 \fi
}%
\providecommand \natexlab [1]{#1}%
\providecommand \enquote  [1]{``#1''}%
\providecommand \bibnamefont  [1]{#1}%
\providecommand \bibfnamefont [1]{#1}%
\providecommand \citenamefont [1]{#1}%
\providecommand \href@noop [0]{\@secondoftwo}%
\providecommand \href [0]{\begingroup \@sanitize@url \@href}%
\providecommand \@href[1]{\@@startlink{#1}\@@href}%
\providecommand \@@href[1]{\endgroup#1\@@endlink}%
\providecommand \@sanitize@url [0]{\catcode `\\12\catcode `\$12\catcode
  `\&12\catcode `\#12\catcode `\^12\catcode `\_12\catcode `\%12\relax}%
\providecommand \@@startlink[1]{}%
\providecommand \@@endlink[0]{}%
\providecommand \url  [0]{\begingroup\@sanitize@url \@url }%
\providecommand \@url [1]{\endgroup\@href {#1}{\urlprefix }}%
\providecommand \urlprefix  [0]{URL }%
\providecommand \Eprint [0]{\href }%
\providecommand \doibase [0]{https://doi.org/}%
\providecommand \selectlanguage [0]{\@gobble}%
\providecommand \bibinfo  [0]{\@secondoftwo}%
\providecommand \bibfield  [0]{\@secondoftwo}%
\providecommand \translation [1]{[#1]}%
\providecommand \BibitemOpen [0]{}%
\providecommand \bibitemStop [0]{}%
\providecommand \bibitemNoStop [0]{.\EOS\space}%
\providecommand \EOS [0]{\spacefactor3000\relax}%
\providecommand \BibitemShut  [1]{\csname bibitem#1\endcsname}%
\let\auto@bib@innerbib\@empty
\bibitem [{\citenamefont {Purcell}(1946)}]{Purcell1946}%
  \BibitemOpen
  \bibfield  {author} {\bibinfo {author} {\bibfnamefont {E.~M.}\ \bibnamefont
  {Purcell}},\ }\bibfield  {title} {\bibinfo {title} {{S}pontaneous emission
  probabilities at radio frequencies},\ }\href@noop {} {\bibfield  {journal}
  {\bibinfo  {journal} {Phys. Rev.}\ }\textbf {\bibinfo {volume} {69}},\
  \bibinfo {pages} {681} (\bibinfo {year} {1946})}\BibitemShut {NoStop}%
\bibitem [{\citenamefont {Koenderink}(2010)}]{Koenderink2010}%
  \BibitemOpen
  \bibfield  {author} {\bibinfo {author} {\bibfnamefont {A.~F.}\ \bibnamefont
  {Koenderink}},\ }\bibfield  {title} {\bibinfo {title} {On the use of purcell
  factors for plasmon antennas},\ }\href {https://doi.org/10.1364/OL.35.004208}
  {\bibfield  {journal} {\bibinfo  {journal} {Opt. Lett.}\ }\textbf {\bibinfo
  {volume} {35}},\ \bibinfo {pages} {4208} (\bibinfo {year}
  {2010})}\BibitemShut {NoStop}%
\bibitem [{\citenamefont {Zambrana-Puyalto}\ and\ \citenamefont
  {Bonod}(2015)}]{Zambrana2015}%
  \BibitemOpen
  \bibfield  {author} {\bibinfo {author} {\bibfnamefont {X.}~\bibnamefont
  {Zambrana-Puyalto}}\ and\ \bibinfo {author} {\bibfnamefont {N.}~\bibnamefont
  {Bonod}},\ }\bibfield  {title} {\bibinfo {title} {Purcell factor of spherical
  mie resonators},\ }\href {https://doi.org/10.1103/PhysRevB.91.195422}
  {\bibfield  {journal} {\bibinfo  {journal} {Phys. Rev. B}\ }\textbf {\bibinfo
  {volume} {91}},\ \bibinfo {pages} {195422} (\bibinfo {year}
  {2015})}\BibitemShut {NoStop}%
\bibitem [{\citenamefont {Novotny}\ and\ \citenamefont
  {Hecht}(2012)}]{novotny}%
  \BibitemOpen
  \bibfield  {author} {\bibinfo {author} {\bibfnamefont {L.}~\bibnamefont
  {Novotny}}\ and\ \bibinfo {author} {\bibfnamefont {B.}~\bibnamefont
  {Hecht}},\ }\href@noop {} {\emph {\bibinfo {title} {{Principles of
  nano-optics. $\mathrm{2nd\, Ed.}$}}}}\ (\bibinfo  {publisher} {New York:
  Cambridge University Press},\ \bibinfo {year} {2012})\BibitemShut {NoStop}%
\bibitem [{\citenamefont {Sprik}\ \emph {et~al.}(1996)\citenamefont {Sprik},
  \citenamefont {van Tiggelen},\ and\ \citenamefont {Lagendijk}}]{sprik_1996}%
  \BibitemOpen
  \bibfield  {author} {\bibinfo {author} {\bibfnamefont {R.}~\bibnamefont
  {Sprik}}, \bibinfo {author} {\bibfnamefont {B.~A.}\ \bibnamefont {van
  Tiggelen}},\ and\ \bibinfo {author} {\bibfnamefont {A.}~\bibnamefont
  {Lagendijk}},\ }\bibfield  {title} {\bibinfo {title} {Optical emission in
  periodic dielectrics},\ }\href@noop {} {\bibfield  {journal} {\bibinfo
  {journal} {Europhys. Lett.}\ }\textbf {\bibinfo {volume} {35}},\ \bibinfo
  {pages} {265} (\bibinfo {year} {1996})}\BibitemShut {NoStop}%
\bibitem [{\citenamefont {Choy}\ \emph {et~al.}(2011)\citenamefont {Choy},
  \citenamefont {Hausmann}, \citenamefont {Babinec}, \citenamefont {Bulu},
  \citenamefont {Khan}, \citenamefont {Maletinsky}, \citenamefont {Yacoby},\
  and\ \citenamefont {Lon\^{c}ar}}]{choy}%
  \BibitemOpen
  \bibfield  {author} {\bibinfo {author} {\bibfnamefont {J.~T.}\ \bibnamefont
  {Choy}}, \bibinfo {author} {\bibfnamefont {B.~J.~M.}\ \bibnamefont
  {Hausmann}}, \bibinfo {author} {\bibfnamefont {T.~M.}\ \bibnamefont
  {Babinec}}, \bibinfo {author} {\bibfnamefont {I.}~\bibnamefont {Bulu}},
  \bibinfo {author} {\bibfnamefont {M.}~\bibnamefont {Khan}}, \bibinfo {author}
  {\bibfnamefont {P.}~\bibnamefont {Maletinsky}}, \bibinfo {author}
  {\bibfnamefont {A.}~\bibnamefont {Yacoby}},\ and\ \bibinfo {author}
  {\bibfnamefont {M.}~\bibnamefont {Lon\^{c}ar}},\ }\bibfield  {title}
  {\bibinfo {title} {Enhanced single-photon emission from a diamond–silver
  aperture},\ }\href@noop {} {\bibfield  {journal} {\bibinfo  {journal} {Nat.
  Photon.}\ }\textbf {\bibinfo {volume} {5}},\ \bibinfo {pages} {738} (\bibinfo
  {year} {2011})}\BibitemShut {NoStop}%
\bibitem [{\citenamefont {Choi}\ \emph {et~al.}(2017)\citenamefont {Choi},
  \citenamefont {Heuck},\ and\ \citenamefont {Englund}}]{PRL_cavity}%
  \BibitemOpen
  \bibfield  {author} {\bibinfo {author} {\bibfnamefont {H.}~\bibnamefont
  {Choi}}, \bibinfo {author} {\bibfnamefont {M.}~\bibnamefont {Heuck}},\ and\
  \bibinfo {author} {\bibfnamefont {D.}~\bibnamefont {Englund}},\ }\bibfield
  {title} {\bibinfo {title} {Self-similar nanocavity design with ultrasmall
  mode volume for single-photon nonlinearities},\ }\href@noop {} {\bibfield
  {journal} {\bibinfo  {journal} {Phys. Rev. Lett.}\ }\textbf {\bibinfo
  {volume} {118}},\ \bibinfo {pages} {223605} (\bibinfo {year}
  {2017})}\BibitemShut {NoStop}%
\bibitem [{\citenamefont {Hu}\ \emph {et~al.}(2018)\citenamefont {Hu},
  \citenamefont {Khater}, \citenamefont {Salas-Montiel}, \citenamefont
  {Kratschmer}, \citenamefont {Engelmann}, \citenamefont {Green},\ and\
  \citenamefont {Weiss}}]{science_cavity}%
  \BibitemOpen
  \bibfield  {author} {\bibinfo {author} {\bibfnamefont {S.}~\bibnamefont
  {Hu}}, \bibinfo {author} {\bibfnamefont {M.}~\bibnamefont {Khater}}, \bibinfo
  {author} {\bibfnamefont {R.}~\bibnamefont {Salas-Montiel}}, \bibinfo {author}
  {\bibfnamefont {E.}~\bibnamefont {Kratschmer}}, \bibinfo {author}
  {\bibfnamefont {S.}~\bibnamefont {Engelmann}}, \bibinfo {author}
  {\bibfnamefont {W.~M.~J.}\ \bibnamefont {Green}},\ and\ \bibinfo {author}
  {\bibfnamefont {S.~M.}\ \bibnamefont {Weiss}},\ }\bibfield  {title} {\bibinfo
  {title} {Experimental realization of deep-subwavelength confinement in
  dielectric optical resonators},\ }\href@noop {} {\bibfield  {journal}
  {\bibinfo  {journal} {Sci. Adv.}\ }\textbf {\bibinfo {volume} {4}},\ \bibinfo
  {pages} {eaat2355} (\bibinfo {year} {2018})}\BibitemShut {NoStop}%
\bibitem [{\citenamefont {Seidler}\ \emph {et~al.}(2013)\citenamefont
  {Seidler}, \citenamefont {Lister}, \citenamefont {Drechsler}, \citenamefont
  {Hofrichter},\ and\ \citenamefont {St\"{o}ferle}}]{Seidler}%
  \BibitemOpen
  \bibfield  {author} {\bibinfo {author} {\bibfnamefont {P.}~\bibnamefont
  {Seidler}}, \bibinfo {author} {\bibfnamefont {K.}~\bibnamefont {Lister}},
  \bibinfo {author} {\bibfnamefont {U.}~\bibnamefont {Drechsler}}, \bibinfo
  {author} {\bibfnamefont {J.}~\bibnamefont {Hofrichter}},\ and\ \bibinfo
  {author} {\bibfnamefont {T.}~\bibnamefont {St\"{o}ferle}},\ }\bibfield
  {title} {\bibinfo {title} {Slotted photonic crystal nanobeam cavity with an
  ultrahigh quality factor-to-mode volume ratio},\ }\href@noop {} {\bibfield
  {journal} {\bibinfo  {journal} {Opt. Express}\ }\textbf {\bibinfo {volume}
  {21}},\ \bibinfo {pages} {32468} (\bibinfo {year} {2013})}\BibitemShut
  {NoStop}%
\bibitem [{\citenamefont {Maier}(2007)}]{Maier:2007}%
  \BibitemOpen
  \bibfield  {author} {\bibinfo {author} {\bibfnamefont {S.~A.}\ \bibnamefont
  {Maier}},\ }\href@noop {} {\emph {\bibinfo {title} {Plasmonics: Fundamentals
  and Applications}}}\ (\bibinfo  {publisher} {Springer},\ \bibinfo {address}
  {New York},\ \bibinfo {year} {2007})\BibitemShut {NoStop}%
\bibitem [{\citenamefont {Doeleman}\ \emph {et~al.}(2016)\citenamefont
  {Doeleman}, \citenamefont {Verhagen},\ and\ \citenamefont
  {Koenderink}}]{Doeleman2016}%
  \BibitemOpen
  \bibfield  {author} {\bibinfo {author} {\bibfnamefont {H.~M.}\ \bibnamefont
  {Doeleman}}, \bibinfo {author} {\bibfnamefont {E.}~\bibnamefont {Verhagen}},\
  and\ \bibinfo {author} {\bibfnamefont {A.~F.}\ \bibnamefont {Koenderink}},\
  }\bibfield  {title} {\bibinfo {title} {{A}ntenna--{C}avity {H}ybrids:
  {M}atching {P}olar {O}pposites for {P}urcell {E}nhancements at any
  {L}inewidth},\ }\href@noop {} {\bibfield  {journal} {\bibinfo  {journal}
  {{A}{C}{S} {P}hotonics}\ }\textbf {\bibinfo {volume} {3}},\ \bibinfo {pages}
  {1943} (\bibinfo {year} {2016})}\BibitemShut {NoStop}%
\bibitem [{\citenamefont {Akselrod}\ \emph {et~al.}(2014)\citenamefont
  {Akselrod}, \citenamefont {Argyropoulos}, \citenamefont {Hoang},
  \citenamefont {Cirac\`i}, \citenamefont {Fang}, \citenamefont {Huang},
  \citenamefont {Smith},\ and\ \citenamefont {Mikkelsen}}]{Akselrod}%
  \BibitemOpen
  \bibfield  {author} {\bibinfo {author} {\bibfnamefont {G.~M.}\ \bibnamefont
  {Akselrod}}, \bibinfo {author} {\bibfnamefont {C.}~\bibnamefont
  {Argyropoulos}}, \bibinfo {author} {\bibfnamefont {T.~B.}\ \bibnamefont
  {Hoang}}, \bibinfo {author} {\bibfnamefont {C.}~\bibnamefont {Cirac\`i}},
  \bibinfo {author} {\bibfnamefont {C.}~\bibnamefont {Fang}}, \bibinfo {author}
  {\bibfnamefont {J.}~\bibnamefont {Huang}}, \bibinfo {author} {\bibfnamefont
  {D.~R.}\ \bibnamefont {Smith}},\ and\ \bibinfo {author} {\bibfnamefont
  {M.~H.}\ \bibnamefont {Mikkelsen}},\ }\bibfield  {title} {\bibinfo {title}
  {Probing the mechanisms of large purcell enhancement in plasmonic
  nanoantennas},\ }\href@noop {} {\bibfield  {journal} {\bibinfo  {journal}
  {Nat. Photon.}\ }\textbf {\bibinfo {volume} {8}},\ \bibinfo {pages} {835}
  (\bibinfo {year} {2014})}\BibitemShut {NoStop}%
\bibitem [{\citenamefont {Mukherjee}\ \emph {et~al.}(2011)\citenamefont
  {Mukherjee}, \citenamefont {Hajisalem},\ and\ \citenamefont
  {Gordon}}]{Mukherjee11}%
  \BibitemOpen
  \bibfield  {author} {\bibinfo {author} {\bibfnamefont {I.}~\bibnamefont
  {Mukherjee}}, \bibinfo {author} {\bibfnamefont {G.}~\bibnamefont
  {Hajisalem}},\ and\ \bibinfo {author} {\bibfnamefont {R.}~\bibnamefont
  {Gordon}},\ }\bibfield  {title} {\bibinfo {title} {One-step integration of
  metal nanoparticle in photonic crystal nanobeam cavity},\ }\href
  {https://doi.org/10.1364/OE.19.022462} {\bibfield  {journal} {\bibinfo
  {journal} {Opt. Express}\ }\textbf {\bibinfo {volume} {19}},\ \bibinfo
  {pages} {22462} (\bibinfo {year} {2011})}\BibitemShut {NoStop}%
\bibitem [{\citenamefont {Kamandar~Dezfouli}\ \emph {et~al.}(2017)\citenamefont
  {Kamandar~Dezfouli}, \citenamefont {Gordon},\ and\ \citenamefont
  {Hughes}}]{PhysRevA.95.013846}%
  \BibitemOpen
  \bibfield  {author} {\bibinfo {author} {\bibfnamefont {M.}~\bibnamefont
  {Kamandar~Dezfouli}}, \bibinfo {author} {\bibfnamefont {R.}~\bibnamefont
  {Gordon}},\ and\ \bibinfo {author} {\bibfnamefont {S.}~\bibnamefont
  {Hughes}},\ }\bibfield  {title} {\bibinfo {title} {Modal theory of modified
  spontaneous emission of a quantum emitter in a hybrid plasmonic
  photonic-crystal cavity system},\ }\href
  {https://doi.org/10.1103/PhysRevA.95.013846} {\bibfield  {journal} {\bibinfo
  {journal} {Phys. Rev. A}\ }\textbf {\bibinfo {volume} {95}},\ \bibinfo
  {pages} {013846} (\bibinfo {year} {2017})}\BibitemShut {NoStop}%
\bibitem [{\citenamefont {Mukherjee}\ and\ \citenamefont
  {Gordon}(2012)}]{Mukherjee12}%
  \BibitemOpen
  \bibfield  {author} {\bibinfo {author} {\bibfnamefont {I.}~\bibnamefont
  {Mukherjee}}\ and\ \bibinfo {author} {\bibfnamefont {R.}~\bibnamefont
  {Gordon}},\ }\bibfield  {title} {\bibinfo {title} {Analysis of hybrid
  plasmonic-photonic crystal structures using perturbation theory},\ }\href
  {https://doi.org/10.1364/OE.20.016992} {\bibfield  {journal} {\bibinfo
  {journal} {Opt. Express}\ }\textbf {\bibinfo {volume} {20}},\ \bibinfo
  {pages} {16992} (\bibinfo {year} {2012})}\BibitemShut {NoStop}%
\bibitem [{\citenamefont {Dezfouli}\ \emph {et~al.}(2019)\citenamefont
  {Dezfouli}, \citenamefont {Gordon},\ and\ \citenamefont {Hughes}}]{Dezfouli}%
  \BibitemOpen
  \bibfield  {author} {\bibinfo {author} {\bibfnamefont {M.~K.}\ \bibnamefont
  {Dezfouli}}, \bibinfo {author} {\bibfnamefont {R.}~\bibnamefont {Gordon}},\
  and\ \bibinfo {author} {\bibfnamefont {S.}~\bibnamefont {Hughes}},\
  }\bibfield  {title} {\bibinfo {title} {{M}olecular {O}ptomechanics in the
  {A}nharmonic {C}avity-{Q}{E}{D} {R}egime {U}sing {H}ybrid
  {M}etal–{D}ielectric {C}avity {M}odes},\ }\href@noop {} {\bibfield
  {journal} {\bibinfo  {journal} {ACS Photonics}\ }\textbf {\bibinfo {volume}
  {6}},\ \bibinfo {pages} {1400} (\bibinfo {year} {2019})}\BibitemShut
  {NoStop}%
\bibitem [{\citenamefont {Palstra}\ \emph {et~al.}(2019)\citenamefont
  {Palstra}, \citenamefont {Doeleman},\ and\ \citenamefont
  {Koenderink}}]{Palstra2019}%
  \BibitemOpen
  \bibfield  {author} {\bibinfo {author} {\bibfnamefont {I.~M.}\ \bibnamefont
  {Palstra}}, \bibinfo {author} {\bibfnamefont {H.~M.}\ \bibnamefont
  {Doeleman}},\ and\ \bibinfo {author} {\bibfnamefont {A.~F.}\ \bibnamefont
  {Koenderink}},\ }\bibfield  {title} {\bibinfo {title} {{H}ybrid
  cavity-antenna systems for quantum optics outside the cryostat?},\
  }\href@noop {} {\bibfield  {journal} {\bibinfo  {journal} {Nanophotonics}\
  }\textbf {\bibinfo {volume} {8}},\ \bibinfo {pages} {1513} (\bibinfo {year}
  {2019})}\BibitemShut {NoStop}%
\bibitem [{\citenamefont {Doeleman}\ \emph {et~al.}(2020)\citenamefont
  {Doeleman}, \citenamefont {Dieleman}, \citenamefont {Mennes}, \citenamefont
  {Ehler},\ and\ \citenamefont {Koenderink}}]{Doeleman2020}%
  \BibitemOpen
  \bibfield  {author} {\bibinfo {author} {\bibfnamefont {H.~M.}\ \bibnamefont
  {Doeleman}}, \bibinfo {author} {\bibfnamefont {C.~D.}\ \bibnamefont
  {Dieleman}}, \bibinfo {author} {\bibfnamefont {C.}~\bibnamefont {Mennes}},
  \bibinfo {author} {\bibfnamefont {B.}~\bibnamefont {Ehler}},\ and\ \bibinfo
  {author} {\bibfnamefont {A.~F.}\ \bibnamefont {Koenderink}},\ }\bibfield
  {title} {\bibinfo {title} {{O}bservation of {C}ooperative {P}urcell
  {E}nhancements in {A}ntenna-{C}avity {H}ybrids},\ }\href@noop {} {\bibfield
  {journal} {\bibinfo  {journal} {{A}{C}{S} Nano}\ }\textbf {\bibinfo {volume}
  {14}},\ \bibinfo {pages} {12027} (\bibinfo {year} {2020})}\BibitemShut
  {NoStop}%
\bibitem [{\citenamefont {Zhang}\ \emph {et~al.}(2020)\citenamefont {Zhang},
  \citenamefont {Liu}, \citenamefont {Wang}, \citenamefont {Zhang},\ and\
  \citenamefont {Lu}}]{Zhang:20}%
  \BibitemOpen
  \bibfield  {author} {\bibinfo {author} {\bibfnamefont {H.}~\bibnamefont
  {Zhang}}, \bibinfo {author} {\bibfnamefont {Y.-C.}\ \bibnamefont {Liu}},
  \bibinfo {author} {\bibfnamefont {C.}~\bibnamefont {Wang}}, \bibinfo {author}
  {\bibfnamefont {N.}~\bibnamefont {Zhang}},\ and\ \bibinfo {author}
  {\bibfnamefont {C.}~\bibnamefont {Lu}},\ }\bibfield  {title} {\bibinfo
  {title} {Hybrid photonic-plasmonic nano-cavity with ultra-high ${Q}/{V}$},\
  }\href@noop {} {\bibfield  {journal} {\bibinfo  {journal} {Opt. Lett.}\
  }\textbf {\bibinfo {volume} {45}},\ \bibinfo {pages} {4794} (\bibinfo {year}
  {2020})}\BibitemShut {NoStop}%
\bibitem [{\citenamefont {Arbabi}\ \emph {et~al.}(2014)\citenamefont {Arbabi},
  \citenamefont {Kamali}, \citenamefont {Arnold},\ and\ \citenamefont
  {Goddard}}]{Ehsan}%
  \BibitemOpen
  \bibfield  {author} {\bibinfo {author} {\bibfnamefont {E.}~\bibnamefont
  {Arbabi}}, \bibinfo {author} {\bibfnamefont {S.~M.}\ \bibnamefont {Kamali}},
  \bibinfo {author} {\bibfnamefont {S.}~\bibnamefont {Arnold}},\ and\ \bibinfo
  {author} {\bibfnamefont {L.~L.}\ \bibnamefont {Goddard}},\ }\bibfield
  {title} {\bibinfo {title} {Hybrid whispering gallery mode/plasmonic chain
  ring resonators for biosensing},\ }\href@noop {} {\bibfield  {journal}
  {\bibinfo  {journal} {Appl. Phys. Lett.}\ }\textbf {\bibinfo {volume}
  {105}},\ \bibinfo {pages} {231107} (\bibinfo {year} {2014})}\BibitemShut
  {NoStop}%
\bibitem [{\citenamefont {Hong}\ \emph {et~al.}(2015)\citenamefont {Hong},
  \citenamefont {Ahn}, \citenamefont {Boriskina}, \citenamefont {Zhao},\ and\
  \citenamefont {Reinhard}}]{Hong}%
  \BibitemOpen
  \bibfield  {author} {\bibinfo {author} {\bibfnamefont {Y.}~\bibnamefont
  {Hong}}, \bibinfo {author} {\bibfnamefont {W.}~\bibnamefont {Ahn}}, \bibinfo
  {author} {\bibfnamefont {S.~V.}\ \bibnamefont {Boriskina}}, \bibinfo {author}
  {\bibfnamefont {X.}~\bibnamefont {Zhao}},\ and\ \bibinfo {author}
  {\bibfnamefont {B.~M.}\ \bibnamefont {Reinhard}},\ }\bibfield  {title}
  {\bibinfo {title} {Directed assembly of optoplasmonic hybrid materials with
  tunable photonic–plasmonic properties},\ }\href@noop {} {\bibfield
  {journal} {\bibinfo  {journal} {J. Phys. Chem. Lett.}\ }\textbf {\bibinfo
  {volume} {6}},\ \bibinfo {pages} {2056} (\bibinfo {year} {2015})}\BibitemShut
  {NoStop}%
\bibitem [{\citenamefont {Klusmann}\ \emph {et~al.}(2017)\citenamefont
  {Klusmann}, \citenamefont {Suryadharma}, \citenamefont {Oppermann},
  \citenamefont {Rockstuhl},\ and\ \citenamefont {Kalt}}]{Klusmann}%
  \BibitemOpen
  \bibfield  {author} {\bibinfo {author} {\bibfnamefont {C.}~\bibnamefont
  {Klusmann}}, \bibinfo {author} {\bibfnamefont {R.~N.~S.}\ \bibnamefont
  {Suryadharma}}, \bibinfo {author} {\bibfnamefont {J.}~\bibnamefont
  {Oppermann}}, \bibinfo {author} {\bibfnamefont {C.}~\bibnamefont
  {Rockstuhl}},\ and\ \bibinfo {author} {\bibfnamefont {H.}~\bibnamefont
  {Kalt}},\ }\bibfield  {title} {\bibinfo {title} {Hybridizing whispering
  gallery modes and plasmonic resonances in a photonic metadevice for
  biosensing applications},\ }\href@noop {} {\bibfield  {journal} {\bibinfo
  {journal} {J. Opt. Soc. Am. B}\ }\textbf {\bibinfo {volume} {34}},\ \bibinfo
  {pages} {D46} (\bibinfo {year} {2017})}\BibitemShut {NoStop}%
\bibitem [{\citenamefont {Hatab}\ \emph {et~al.}(2010)\citenamefont {Hatab},
  \citenamefont {Hsueh}, \citenamefont {Gaddis}, \citenamefont {Retterer},
  \citenamefont {Li}, \citenamefont {Eres}, \citenamefont {Zhang},\ and\
  \citenamefont {Gu}}]{Hatab2010}%
  \BibitemOpen
  \bibfield  {author} {\bibinfo {author} {\bibfnamefont {N.~A.}\ \bibnamefont
  {Hatab}}, \bibinfo {author} {\bibfnamefont {C.-H.}\ \bibnamefont {Hsueh}},
  \bibinfo {author} {\bibfnamefont {A.~L.}\ \bibnamefont {Gaddis}}, \bibinfo
  {author} {\bibfnamefont {S.~T.}\ \bibnamefont {Retterer}}, \bibinfo {author}
  {\bibfnamefont {J.-H.}\ \bibnamefont {Li}}, \bibinfo {author} {\bibfnamefont
  {G.}~\bibnamefont {Eres}}, \bibinfo {author} {\bibfnamefont {Z.}~\bibnamefont
  {Zhang}},\ and\ \bibinfo {author} {\bibfnamefont {B.}~\bibnamefont {Gu}},\
  }\bibfield  {title} {\bibinfo {title} {{F}ree-standing optical gold bowtie
  nanoantenna with variable gap size for enhanced {R}aman spectroscopy},\
  }\href@noop {} {\bibfield  {journal} {\bibinfo  {journal} {Nano Lett.}\
  }\textbf {\bibinfo {volume} {10}},\ \bibinfo {pages} {4952} (\bibinfo {year}
  {2010})}\BibitemShut {NoStop}%
\bibitem [{\citenamefont {Zhu}\ and\ \citenamefont {Crozier}(2014)}]{Crozier}%
  \BibitemOpen
  \bibfield  {author} {\bibinfo {author} {\bibfnamefont {W.}~\bibnamefont
  {Zhu}}\ and\ \bibinfo {author} {\bibfnamefont {K.~B.}\ \bibnamefont
  {Crozier}},\ }\bibfield  {title} {\bibinfo {title} {Quantum mechanical limit
  to plasmonic enhancement as observed by surface-enhanced {R}aman
  scattering},\ }\href@noop {} {\bibfield  {journal} {\bibinfo  {journal} {Nat.
  Commun.}\ }\textbf {\bibinfo {volume} {5}},\ \bibinfo {pages} {5228}
  (\bibinfo {year} {2014})}\BibitemShut {NoStop}%
\bibitem [{\citenamefont {Duan}\ \emph {et~al.}(2012)\citenamefont {Duan},
  \citenamefont {Fern\'{a}ndez-Dom\'{i}nguez}, \citenamefont {Bosman},
  \citenamefont {Maier},\ and\ \citenamefont {Yang}}]{Duan_2012}%
  \BibitemOpen
  \bibfield  {author} {\bibinfo {author} {\bibfnamefont {H.}~\bibnamefont
  {Duan}}, \bibinfo {author} {\bibfnamefont {A.~I.}\ \bibnamefont
  {Fern\'{a}ndez-Dom\'{i}nguez}}, \bibinfo {author} {\bibfnamefont
  {M.}~\bibnamefont {Bosman}}, \bibinfo {author} {\bibfnamefont {S.~A.}\
  \bibnamefont {Maier}},\ and\ \bibinfo {author} {\bibfnamefont {J.~K.~W.}\
  \bibnamefont {Yang}},\ }\bibfield  {title} {\bibinfo {title} {Nanoplasmonics:
  {C}lassical down to the {N}anometer {S}cale},\ }\href@noop {} {\bibfield
  {journal} {\bibinfo  {journal} {Nano Lett.}\ }\textbf {\bibinfo {volume}
  {12}},\ \bibinfo {pages} {1683} (\bibinfo {year} {2012})}\BibitemShut
  {NoStop}%
\bibitem [{\citenamefont {L\'{e}v\^{e}que}\ and\ \citenamefont
  {Martin}(2006)}]{Leveque2006}%
  \BibitemOpen
  \bibfield  {author} {\bibinfo {author} {\bibfnamefont {G.}~\bibnamefont
  {L\'{e}v\^{e}que}}\ and\ \bibinfo {author} {\bibfnamefont {O.~J.~F.}\
  \bibnamefont {Martin}},\ }\bibfield  {title} {\bibinfo {title} {{O}ptical
  interactions in a plasmonic particle coupled to a metallic film},\
  }\href@noop {} {\bibfield  {journal} {\bibinfo  {journal} {Opt. Express}\
  }\textbf {\bibinfo {volume} {14}},\ \bibinfo {pages} {9971} (\bibinfo {year}
  {2006})}\BibitemShut {NoStop}%
\bibitem [{\citenamefont {Cirac\`{i}}\ \emph {et~al.}(2012)\citenamefont
  {Cirac\`{i}}, \citenamefont {Hill}, \citenamefont {Mock}, \citenamefont
  {Urzhumov}, \citenamefont {Fern\'{a}ndez-Dom\'{i}nguez}, \citenamefont
  {Maier}, \citenamefont {Pendry}, \citenamefont {Chilkoti},\ and\
  \citenamefont {Smith}}]{Ciraci2012}%
  \BibitemOpen
  \bibfield  {author} {\bibinfo {author} {\bibfnamefont {C.}~\bibnamefont
  {Cirac\`{i}}}, \bibinfo {author} {\bibfnamefont {R.~T.}\ \bibnamefont
  {Hill}}, \bibinfo {author} {\bibfnamefont {J.~J.}\ \bibnamefont {Mock}},
  \bibinfo {author} {\bibfnamefont {Y.}~\bibnamefont {Urzhumov}}, \bibinfo
  {author} {\bibfnamefont {A.~I.}\ \bibnamefont {Fern\'{a}ndez-Dom\'{i}nguez}},
  \bibinfo {author} {\bibfnamefont {S.~A.}\ \bibnamefont {Maier}}, \bibinfo
  {author} {\bibfnamefont {J.~B.}\ \bibnamefont {Pendry}}, \bibinfo {author}
  {\bibfnamefont {A.}~\bibnamefont {Chilkoti}},\ and\ \bibinfo {author}
  {\bibfnamefont {D.~R.}\ \bibnamefont {Smith}},\ }\bibfield  {title} {\bibinfo
  {title} {{P}robing the ultimate limits of plasmonic enhancement},\
  }\href@noop {} {\bibfield  {journal} {\bibinfo  {journal} {Science}\ }\textbf
  {\bibinfo {volume} {337}},\ \bibinfo {pages} {1072} (\bibinfo {year}
  {2012})}\BibitemShut {NoStop}%
\bibitem [{\citenamefont {Baumberg}\ \emph {et~al.}(2019)\citenamefont
  {Baumberg}, \citenamefont {Aizpurua}, \citenamefont {Mikkelsen},\ and\
  \citenamefont {Smith}}]{Baumberg2019}%
  \BibitemOpen
  \bibfield  {author} {\bibinfo {author} {\bibfnamefont {J.~J.}\ \bibnamefont
  {Baumberg}}, \bibinfo {author} {\bibfnamefont {J.}~\bibnamefont {Aizpurua}},
  \bibinfo {author} {\bibfnamefont {M.~H.}\ \bibnamefont {Mikkelsen}},\ and\
  \bibinfo {author} {\bibfnamefont {D.~R.}\ \bibnamefont {Smith}},\ }\bibfield
  {title} {\bibinfo {title} {{E}xtreme nanophotonics from ultrathin metallic
  gaps},\ }\href@noop {} {\bibfield  {journal} {\bibinfo  {journal} {Nat.
  Mater.}\ }\textbf {\bibinfo {volume} {18}},\ \bibinfo {pages} {668} (\bibinfo
  {year} {2019})}\BibitemShut {NoStop}%
\bibitem [{\citenamefont {Carnegie}\ \emph {et~al.}(2018)\citenamefont
  {Carnegie}, \citenamefont {Griffiths}, \citenamefont {de~Nijs}, \citenamefont
  {Readman}, \citenamefont {Chikkaraddy}, \citenamefont {Deacon}, \citenamefont
  {Zhang}, \citenamefont {Szab\'{o}}, \citenamefont {Rosta}, \citenamefont
  {Aizpurua},\ and\ \citenamefont {Baumberg}}]{Carnegie2018}%
  \BibitemOpen
  \bibfield  {author} {\bibinfo {author} {\bibfnamefont {C.}~\bibnamefont
  {Carnegie}}, \bibinfo {author} {\bibfnamefont {J.}~\bibnamefont {Griffiths}},
  \bibinfo {author} {\bibfnamefont {B.}~\bibnamefont {de~Nijs}}, \bibinfo
  {author} {\bibfnamefont {C.}~\bibnamefont {Readman}}, \bibinfo {author}
  {\bibfnamefont {R.}~\bibnamefont {Chikkaraddy}}, \bibinfo {author}
  {\bibfnamefont {W.~M.}\ \bibnamefont {Deacon}}, \bibinfo {author}
  {\bibfnamefont {Y.}~\bibnamefont {Zhang}}, \bibinfo {author} {\bibfnamefont
  {I.}~\bibnamefont {Szab\'{o}}}, \bibinfo {author} {\bibfnamefont
  {E.}~\bibnamefont {Rosta}}, \bibinfo {author} {\bibfnamefont
  {J.}~\bibnamefont {Aizpurua}},\ and\ \bibinfo {author} {\bibfnamefont
  {J.~J.}\ \bibnamefont {Baumberg}},\ }\bibfield  {title} {\bibinfo {title}
  {{R}oom-{T}emperature {O}ptical {P}icocavities below 1 nm$^{3}$ {A}ccessing
  {S}ingle-{A}tom {G}eometries},\ }\href@noop {} {\bibfield  {journal}
  {\bibinfo  {journal} {J. Phys. Chem. Lett.}\ }\textbf {\bibinfo {volume}
  {9}},\ \bibinfo {pages} {7146} (\bibinfo {year} {2018})}\BibitemShut
  {NoStop}%
\bibitem [{\citenamefont {Barbry}\ \emph {et~al.}(2015)\citenamefont {Barbry},
  \citenamefont {Koval}, \citenamefont {Marchesin}, \citenamefont {Esteban},
  \citenamefont {Borisov}, \citenamefont {Aizpurua},\ and\ \citenamefont
  {S\'{a}nchez-Portal}}]{Barbry}%
  \BibitemOpen
  \bibfield  {author} {\bibinfo {author} {\bibfnamefont {M.}~\bibnamefont
  {Barbry}}, \bibinfo {author} {\bibfnamefont {P.}~\bibnamefont {Koval}},
  \bibinfo {author} {\bibfnamefont {F.}~\bibnamefont {Marchesin}}, \bibinfo
  {author} {\bibfnamefont {R.}~\bibnamefont {Esteban}}, \bibinfo {author}
  {\bibfnamefont {A.~G.}\ \bibnamefont {Borisov}}, \bibinfo {author}
  {\bibfnamefont {J.}~\bibnamefont {Aizpurua}},\ and\ \bibinfo {author}
  {\bibfnamefont {D.}~\bibnamefont {S\'{a}nchez-Portal}},\ }\bibfield  {title}
  {\bibinfo {title} {Atomistic {N}ear-{F}ield {N}anoplasmonics: {R}eaching
  {A}tomic-{S}cale {R}esolution in {N}anooptics},\ }\href@noop {} {\bibfield
  {journal} {\bibinfo  {journal} {Nano Lett.}\ }\textbf {\bibinfo {volume}
  {15}},\ \bibinfo {pages} {3410} (\bibinfo {year} {2015})}\BibitemShut
  {NoStop}%
\bibitem [{\citenamefont {Barreda}\ \emph {et~al.}(2021)\citenamefont
  {Barreda}, \citenamefont {Zapata-Herrera}, \citenamefont {Palstra},
  \citenamefont {Mercad\'{e}}, \citenamefont {Aizpurua}, \citenamefont
  {Koenderink},\ and\ \citenamefont {Mart\'{i}nez}}]{Barreda:21}%
  \BibitemOpen
  \bibfield  {author} {\bibinfo {author} {\bibfnamefont {A.~I.}\ \bibnamefont
  {Barreda}}, \bibinfo {author} {\bibfnamefont {M.}~\bibnamefont
  {Zapata-Herrera}}, \bibinfo {author} {\bibfnamefont {I.~M.}\ \bibnamefont
  {Palstra}}, \bibinfo {author} {\bibfnamefont {L.}~\bibnamefont
  {Mercad\'{e}}}, \bibinfo {author} {\bibfnamefont {J.}~\bibnamefont
  {Aizpurua}}, \bibinfo {author} {\bibfnamefont {A.~F.}\ \bibnamefont
  {Koenderink}},\ and\ \bibinfo {author} {\bibfnamefont {A.}~\bibnamefont
  {Mart\'{i}nez}},\ }\bibfield  {title} {\bibinfo {title} {Hybrid
  photonic-plasmonic cavities based on the nanoparticle-on-a-mirror
  configuration},\ }\href@noop {} {\bibfield  {journal} {\bibinfo  {journal}
  {Photon. Res.}\ }\textbf {\bibinfo {volume} {9}},\ \bibinfo {pages} {2398}
  (\bibinfo {year} {2021})}\BibitemShut {NoStop}%
\bibitem [{\citenamefont {Deotare}\ \emph {et~al.}(2009)\citenamefont
  {Deotare}, \citenamefont {McCutcheon}, \citenamefont {Frank}, \citenamefont
  {Khan},\ and\ \citenamefont {Lončar}}]{apl}%
  \BibitemOpen
  \bibfield  {author} {\bibinfo {author} {\bibfnamefont {P.~B.}\ \bibnamefont
  {Deotare}}, \bibinfo {author} {\bibfnamefont {M.~W.}\ \bibnamefont
  {McCutcheon}}, \bibinfo {author} {\bibfnamefont {I.~W.}\ \bibnamefont
  {Frank}}, \bibinfo {author} {\bibfnamefont {M.}~\bibnamefont {Khan}},\ and\
  \bibinfo {author} {\bibfnamefont {M.}~\bibnamefont {Lončar}},\ }\bibfield
  {title} {\bibinfo {title} {High quality factor photonic crystal nanobeam
  cavities},\ }\href@noop {} {\bibfield  {journal} {\bibinfo  {journal} {Appl.
  Phys. Lett.}\ }\textbf {\bibinfo {volume} {94}},\ \bibinfo {pages} {121106}
  (\bibinfo {year} {2009})}\BibitemShut {NoStop}%
\bibitem [{\citenamefont {Akahane}\ \emph {et~al.}(2003)\citenamefont
  {Akahane}, \citenamefont {Asano}, \citenamefont {Song},\ and\ \citenamefont
  {Noda}}]{nature}%
  \BibitemOpen
  \bibfield  {author} {\bibinfo {author} {\bibfnamefont {Y.}~\bibnamefont
  {Akahane}}, \bibinfo {author} {\bibfnamefont {T.}~\bibnamefont {Asano}},
  \bibinfo {author} {\bibfnamefont {B.-S.}\ \bibnamefont {Song}},\ and\
  \bibinfo {author} {\bibfnamefont {S.}~\bibnamefont {Noda}},\ }\bibfield
  {title} {\bibinfo {title} {High-$q$ photonic nanocavity in a two-dimensional
  photonic crystal},\ }\href@noop {} {\bibfield  {journal} {\bibinfo  {journal}
  {Nature}\ }\textbf {\bibinfo {volume} {425}},\ \bibinfo {pages} {944}
  (\bibinfo {year} {2003})}\BibitemShut {NoStop}%
\bibitem [{\citenamefont {Robinson}\ \emph {et~al.}(2005)\citenamefont
  {Robinson}, \citenamefont {Manolatou}, \citenamefont {Chen},\ and\
  \citenamefont {Lipson}}]{Robinson}%
  \BibitemOpen
  \bibfield  {author} {\bibinfo {author} {\bibfnamefont {J.~T.}\ \bibnamefont
  {Robinson}}, \bibinfo {author} {\bibfnamefont {C.}~\bibnamefont {Manolatou}},
  \bibinfo {author} {\bibfnamefont {L.}~\bibnamefont {Chen}},\ and\ \bibinfo
  {author} {\bibfnamefont {M.}~\bibnamefont {Lipson}},\ }\bibfield  {title}
  {\bibinfo {title} {Ultrasmall mode volumes in dielectric optical
  microcavities},\ }\href@noop {} {\bibfield  {journal} {\bibinfo  {journal}
  {Phys. Rev. Lett.}\ }\textbf {\bibinfo {volume} {95}},\ \bibinfo {pages}
  {143901} (\bibinfo {year} {2005})}\BibitemShut {NoStop}%
\bibitem [{\citenamefont {{COMSOL Multiphysics 5.0}}()}]{Comsol}%
  \BibitemOpen
  \bibfield  {author} {\bibinfo {author} {\bibnamefont {{COMSOL Multiphysics
  5.0}}},\ }\href@noop {} {\bibinfo {title} {{(COMSOL Inc., 2015)}}},\ \bibinfo
  {howpublished} {\url{https://www.comsol.com}}\BibitemShut {NoStop}%
\bibitem [{\citenamefont {Palik}(1998)}]{Palik1998}%
  \BibitemOpen
  \bibfield  {author} {\bibinfo {author} {\bibfnamefont {E.~D.}\ \bibnamefont
  {Palik}},\ }\href@noop {} {\emph {\bibinfo {title} {{Handbook of optical
  constants of solids}}}}\ (\bibinfo  {publisher} {Academic press},\ \bibinfo
  {year} {1998})\BibitemShut {NoStop}%
\bibitem [{\citenamefont {Mercadé}\ \emph {et~al.}(2020)\citenamefont
  {Mercadé}, \citenamefont {Martín}, \citenamefont {Griol}, \citenamefont
  {Navarro-Urrios},\ and\ \citenamefont {Mart\'inez}}]{mercade}%
  \BibitemOpen
  \bibfield  {author} {\bibinfo {author} {\bibfnamefont {L.}~\bibnamefont
  {Mercadé}}, \bibinfo {author} {\bibfnamefont {L.~L.}\ \bibnamefont
  {Martín}}, \bibinfo {author} {\bibfnamefont {A.}~\bibnamefont {Griol}},
  \bibinfo {author} {\bibfnamefont {D.}~\bibnamefont {Navarro-Urrios}},\ and\
  \bibinfo {author} {\bibfnamefont {A.}~\bibnamefont {Mart\'inez}},\ }\bibfield
   {title} {\bibinfo {title} {Microwave oscillator and frequency comb in a
  silicon optomechanical cavity with a full phononic bandgap},\ }\href@noop {}
  {\bibfield  {journal} {\bibinfo  {journal} {Nanophotonics}\ }\textbf
  {\bibinfo {volume} {9}},\ \bibinfo {pages} {3535} (\bibinfo {year}
  {2020})}\BibitemShut {NoStop}%
\bibitem [{\citenamefont {Chen}\ \emph {et~al.}(2021)\citenamefont {Chen},
  \citenamefont {Roelli}, \citenamefont {Hu}, \citenamefont {Verlekar},
  \citenamefont {Amirtharaj}, \citenamefont {Barreda}, \citenamefont
  {Kippenberg}, \citenamefont {Kovylina}, \citenamefont {Verhagen},
  \citenamefont {Mart\'inez},\ and\ \citenamefont {Galland}}]{science_galland}%
  \BibitemOpen
  \bibfield  {author} {\bibinfo {author} {\bibfnamefont {W.}~\bibnamefont
  {Chen}}, \bibinfo {author} {\bibfnamefont {P.}~\bibnamefont {Roelli}},
  \bibinfo {author} {\bibfnamefont {H.}~\bibnamefont {Hu}}, \bibinfo {author}
  {\bibfnamefont {S.}~\bibnamefont {Verlekar}}, \bibinfo {author}
  {\bibfnamefont {S.~P.}\ \bibnamefont {Amirtharaj}}, \bibinfo {author}
  {\bibfnamefont {A.~I.}\ \bibnamefont {Barreda}}, \bibinfo {author}
  {\bibfnamefont {T.~J.}\ \bibnamefont {Kippenberg}}, \bibinfo {author}
  {\bibfnamefont {M.}~\bibnamefont {Kovylina}}, \bibinfo {author}
  {\bibfnamefont {E.}~\bibnamefont {Verhagen}}, \bibinfo {author}
  {\bibfnamefont {A.}~\bibnamefont {Mart\'inez}},\ and\ \bibinfo {author}
  {\bibfnamefont {C.}~\bibnamefont {Galland}},\ }\bibfield  {title} {\bibinfo
  {title} {Continuous-wave frequency upconversion with a molecular
  optomechanical nanocavity},\ }\href@noop {} {\bibfield  {journal} {\bibinfo
  {journal} {Science}\ }\textbf {\bibinfo {volume} {374}},\ \bibinfo {pages}
  {1264} (\bibinfo {year} {2021})}\BibitemShut {NoStop}%
\bibitem [{\citenamefont {Yang}\ \emph {et~al.}(2009)\citenamefont {Yang},
  \citenamefont {Moore}, \citenamefont {Schmidt}, \citenamefont {Klug},
  \citenamefont {Lipson},\ and\ \citenamefont {Erickson}}]{nature_n}%
  \BibitemOpen
  \bibfield  {author} {\bibinfo {author} {\bibfnamefont {A.~H.~J.}\
  \bibnamefont {Yang}}, \bibinfo {author} {\bibfnamefont {S.~D.}\ \bibnamefont
  {Moore}}, \bibinfo {author} {\bibfnamefont {B.~S.}\ \bibnamefont {Schmidt}},
  \bibinfo {author} {\bibfnamefont {M.}~\bibnamefont {Klug}}, \bibinfo {author}
  {\bibfnamefont {M.}~\bibnamefont {Lipson}},\ and\ \bibinfo {author}
  {\bibfnamefont {D.}~\bibnamefont {Erickson}},\ }\bibfield  {title} {\bibinfo
  {title} {Optical manipulation of nanoparticles and biomolecules in
  sub-wavelength slot waveguides},\ }\href@noop {} {\bibfield  {journal}
  {\bibinfo  {journal} {Nature}\ }\textbf {\bibinfo {volume} {457}},\ \bibinfo
  {pages} {71} (\bibinfo {year} {2009})}\BibitemShut {NoStop}%
\bibitem [{\citenamefont {Losada}\ \emph {et~al.}(2019)\citenamefont {Losada},
  \citenamefont {Raza}, \citenamefont {Clemmen}, \citenamefont {Serrano},
  \citenamefont {Griol}, \citenamefont {Baets},\ and\ \citenamefont
  {Martínez}}]{LOS19-JSTQE}%
  \BibitemOpen
  \bibfield  {author} {\bibinfo {author} {\bibfnamefont {J.}~\bibnamefont
  {Losada}}, \bibinfo {author} {\bibfnamefont {A.}~\bibnamefont {Raza}},
  \bibinfo {author} {\bibfnamefont {S.}~\bibnamefont {Clemmen}}, \bibinfo
  {author} {\bibfnamefont {A.}~\bibnamefont {Serrano}}, \bibinfo {author}
  {\bibfnamefont {A.}~\bibnamefont {Griol}}, \bibinfo {author} {\bibfnamefont
  {R.}~\bibnamefont {Baets}},\ and\ \bibinfo {author} {\bibfnamefont
  {A.}~\bibnamefont {Martínez}},\ }\bibfield  {title} {\bibinfo {title} {Sers
  detection via individual bowtie nanoantennas integrated in
  si<sub>3</sub>n<sub>4</sub> waveguides},\ }\href
  {https://doi.org/10.1109/JSTQE.2019.2896200} {\bibfield  {journal} {\bibinfo
  {journal} {IEEE Journal of Selected Topics in Quantum Electronics}\ }\textbf
  {\bibinfo {volume} {25}},\ \bibinfo {pages} {1} (\bibinfo {year}
  {2019})}\BibitemShut {NoStop}%
\bibitem [{\citenamefont {Xomalis}\ \emph {et~al.}(2021)\citenamefont
  {Xomalis}, \citenamefont {Zheng}, \citenamefont {Chikkaraddy}, \citenamefont
  {Koczor-Benda}, \citenamefont {Miele}, \citenamefont {Rosta}, \citenamefont
  {Vandenbosch}, \citenamefont {Martínez},\ and\ \citenamefont
  {Baumberg}}]{science_baumberg}%
  \BibitemOpen
  \bibfield  {author} {\bibinfo {author} {\bibfnamefont {A.}~\bibnamefont
  {Xomalis}}, \bibinfo {author} {\bibfnamefont {X.}~\bibnamefont {Zheng}},
  \bibinfo {author} {\bibfnamefont {R.}~\bibnamefont {Chikkaraddy}}, \bibinfo
  {author} {\bibfnamefont {Z.}~\bibnamefont {Koczor-Benda}}, \bibinfo {author}
  {\bibfnamefont {E.}~\bibnamefont {Miele}}, \bibinfo {author} {\bibfnamefont
  {E.}~\bibnamefont {Rosta}}, \bibinfo {author} {\bibfnamefont {G.~A.~E.}\
  \bibnamefont {Vandenbosch}}, \bibinfo {author} {\bibfnamefont
  {A.}~\bibnamefont {Martínez}},\ and\ \bibinfo {author} {\bibfnamefont
  {J.~J.}\ \bibnamefont {Baumberg}},\ }\bibfield  {title} {\bibinfo {title}
  {Detecting mid-infrared light by molecular frequency upconversion in
  dual-wavelength nanoantennas},\ }\href
  {https://doi.org/10.1126/science.abk2593} {\bibfield  {journal} {\bibinfo
  {journal} {Science}\ }\textbf {\bibinfo {volume} {374}},\ \bibinfo {pages}
  {1268} (\bibinfo {year} {2021})},\ \Eprint
  {https://arxiv.org/abs/https://www.science.org/doi/pdf/10.1126/science.abk2593}
  {https://www.science.org/doi/pdf/10.1126/science.abk2593} \BibitemShut
  {NoStop}%
\end{thebibliography}

%

\end{document}